# "Change" in Conceptual Modeling and Systems Reconfiguration

Sabah Al-Fedaghi
*salfedaghi@yahoo.com, sabah.alfedaghi@ku.edu.kw*
Computer Engineering Department, Kuwait University, Kuwait

**Summary**
In this paper, we explore the notion of change in systems and software engineering, emphasizing its philosophical elucidation. Generally, it has been claimed that change is so pervasive in systems that it almost defeats description and analysis. In this article, we analyze *change* using the conceptual modeling technique called a *thinging machine* (TM), which reflects change in terms of the actions of creating, processing, releasing, transferring, and receiving things. We illustrated change in TM modeling with an example of a system's reconfiguration of business product handling designed using business process modeling notation (BPMN). Then we analyze the notion of change and compare its various definitions in philosophy. Specifically, we examine Zeno's paradox that involves how to account for change and continuity together in moving things. The problem is that we cannot assert that an arrow is actually moving when it has been shot from a bow because the arrow needs to be at a certain place at each point in time, which by definition cannot contain any duration at all. In our analysis of this problem, we convert the arrow trajectory into space units called thimacs. In the TM generic actions, two types of change are identified: *state* and *progression* (PROCESS) changes. Therefore, when an arrow flows to a TM machine that represents a trajectory space unit, it is rejected, causing it to bounce away to the outside. That is, the arrow is transferred, arrives, and is transferred back; therefore, the arrow is never *accepted* into a thimac in the trajectory at any moment. The result of such analysis seems to introduce a logical explanation for the notion of movement discussed in Zeno's puzzles.

## 1. Introduction

In software and systems engineering, adaptation to future user requirements or changing domain-imposed requirements is an essential consideration in the system's development and operating environment [1]. Systems are expected to operate in dynamic environments and to deal with the new problems and arbitrary changes. To cope with these changes, a reconfiguration process may be applied to rearrange the system components concerned with deliberate modifications to technical and organizational subsystems. Frequent changes are made to update hardware and software components, fix software flaws and other errors, address security threats, and adapt to changing business objectives [2].

This paper focuses on schematic changes involving structural changes resulting from altering requirements and bringing a system into compliance with those requirements. Such changes necessitate modifying the conceptual specification of the involved system. System specifications can be developed at various levels of abstraction, with transformations ranging from high- to low-level specifications. A high-level specification describes the overall configuration of a system.

Change is a central concept for software and systems engineering. According to Idris [3], "The meaning of change is one of the fundamental subjects of inquiry of philosophy. It plays a substantial role in providing our understanding of reality." Such issues as what change is, how it happens, and how we can know it has happened are essential to our understanding of systems. Examining the paradoxical nature of change, we can gain new insights into theory and practice in many scientific fields [4].

We further explore the notion of change utilizing our thinging machine (TM) model. Our aim is to develop a better understanding of the notion of change to appreciate the nature of the field of software and systems engineering. Such a field of inquiry, despite considerable technical developments, has yet to form a coherent and theoretical framework for its key notions. Accordingly, we attempt to offer a conceptual framework to study theoretical and trans-disciplinary foundations of the notion of change. According to Uysal [5], "Investigating the trans-disciplinary aspects of [software engineering] may pave the way of some solutions while it may shed light on building theoretical background of possible empirical studies. However, the review of [software engineering] literature shows the little effort given to this research gap."

The next section provides a brief description of the TM model. Section 3 gives a sample TM application in software engineering in the form of a conceptual model of a business orders system. The example involves modification of an ordering conceptual schema as an instance of changes in requirements in the original description of the system. In section 4, we focus on reviewing the notion of change. In section 5, we apply the TM model to change in the philosophical sense with the problem of an arrow's movement in Zeno's paradox.

## 2. Thinging Machine (TM) Modeling

The traditional ontology divides entities into objects, which are extended in space, and processes, which are extended in time. TM modeling introduces a drastically different conceptualization, which consists of a lower (static) characterization of entities as things that are simultaneously machines, and both merge into a complex of interrelated entities called thimacs [6]. At the upper level (dynamics), a time thimac combines with the static thimac to initiate events (See Fig. 1).

The thing and the corresponding machine "exist" as one thimac; the thing reflects the unity, and the machine shows the structural components, including *potential* (static: outside of time) actions of behavior. The static "thing" does not actually exist, change, or move, but it has potentialities for these actions when combined with time. Such a view reminds us of the wave particle dualism of quantum mechanics.

A thimac is a thing. The thing is what can be created (appear, observed), processed (changed), released, transferred, and/or received. As we will discuss later, a thing is manifested (can be recognized as a unity) and related to the whole TM or as a static (timeless) phenomenon. This whole TM occupies a conceptual "space" that forms a network of co-existing thimacs. The whole is a grand thing/machine. Thimacs can be "located" only via flow connections among thimacs. Later, when we discuss dynamism, this thing becomes an "instance" when supplemented with time (which is also a thing) to form a dynamic unity of a thing called an *event*. Therefore, things are part of the TM static description (model) and are part of the dynamic model when merged with time.

The thimac is also a machine that creates, processes, releases, transfers, and/or receives. Fig. 2 shows a general picture of a machine. The figure indicates five "seeds" of potentialities of dynamism: creation, processing, release, transfer, and receive.

All things are created, processed, and transported (acted on), and all machines (thimacs) create, process, and transport other things. Things "flow through" (denoted by a solid arrow in Fig. 2) other machines. The thing in a TM diagram is a presentation of any "existing" (appearing) entity that can be "counted as one" and is coherent as a unity.

Fig. 2 can be described in terms of the following generic (has no more primitive action) actions:
**Arrive**: A thing moves to a machine.
**Accept**: A thing enters the machine. For simplification, we assume that all arriving things are accepted; thus, we can combine the arrive and accept stages into one stage: the **receive** stage.
**Release**: A thing is ready for transfer outside the machine.
**Process**: A thing is changed, handled, and examined, but no new thing results.
**Create**: A new thing is born (found/manifested) in the machine and is realized from the moment it arises (emergence) in a thimac. Things come into being in the model by "being found."
**Transfer**: A thing is input into or output from a machine.

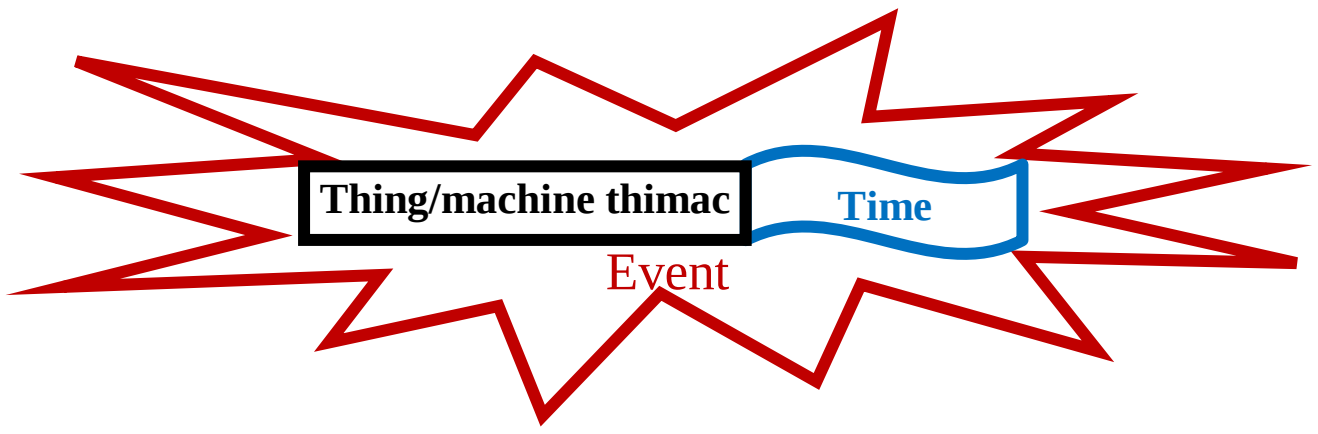

Fig. 1 Static and dynamic thimacs

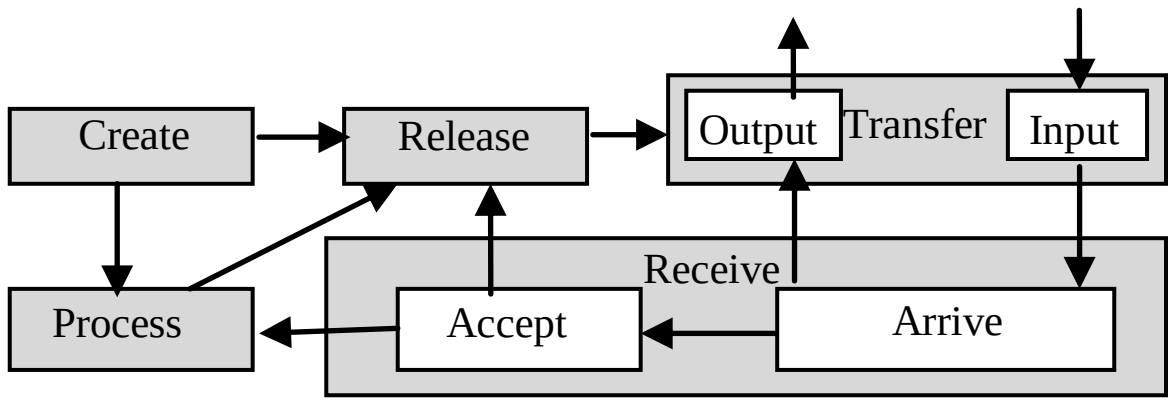

Fig. 2 Thing machine

Additionally, the TM model includes the **triggering** mechanism (denoted by a dashed arrow in this article's figures), which initiates a flow from one machine to another. Multiple machines can interact with each other through the movement of things or through triggering. Triggering is a transformation from one series of movements to another.

## 3. Example of TM Modeling

According to Zhou [7], systems are increasingly expected to operate in dynamic environments to deal with the new problems and tasks, and the system's flexibility requires dynamic *reconfiguration*. "Reconfiguration" refers to changes such as adding actions or deleting components, changing links between systems, and modifying component configuration. Zhou [7] emphasized that systems have to face problems caused by continually developing new web services and modifying or terminating existing web services. The area of service reconfiguration requires more work on the modeling and verification of dynamic reconfiguration of dependable services, including the problem of interface between old configuration activities, new configuration activities, and reconfiguration activities [7].

For example, Zhou [7] designed a system using business process modeling notation (BPMN). This case study was conducted on an organization that handles product orders from customers. When the organization receives an order from a customer, a form is filled out. Then this form is sent to credit check and passes to inventory check. After this evaluation, the order is rejected, or it is processed and passed on to billing and shipping. The billing procedure bills for the total cost of ordered items plus their shipping costs. Afterward, the order is archived and a confirmation notification is sent to the customer. Fig. 3 shows a partial view of the BPMN model of the given case study.

## 3.1 Reconfiguration

According to Zhou [7], the company decides to reconfigure the order of billing and shipping activities, moving the billing activity to occur before the shipping activity and keeping the old configuration process and new configuration process simultaneously available. This reconfiguration requires change only in the system's main lane while the rest of the workflow remains the same. Fig. 4 shows the modified part of the new configuration's BPMN diagram. Comparing this BPMN to the old BPMN diagram in Fig. 3, the two parallel gateways in the main lane have been removed, and the two activities are now called synchronously to keep the old and new configuration processes simultaneously available. Accordingly, Zhou [7] defines a default flow that is identical to the old configuration. This default flow can be altered by an interrupting message event contained in the "determine configuration" activity included in a separate reconfigure region pool. This activity determines which configuration should be used when the system is called. The modified part of new configuration BPMN diagram is shown in Fig. 4.

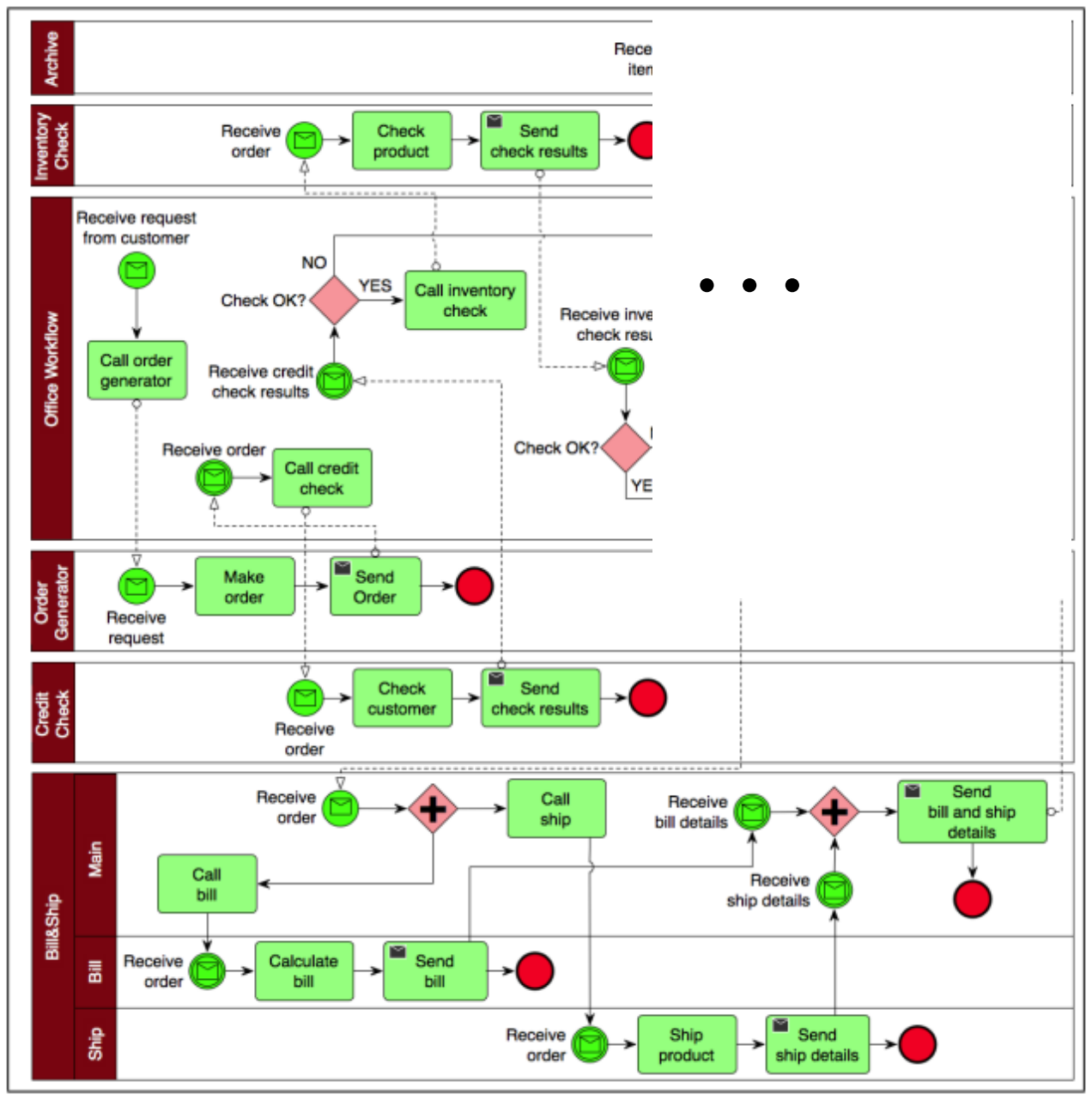

Fig. 3 Case Study Workflow - BPMN Model

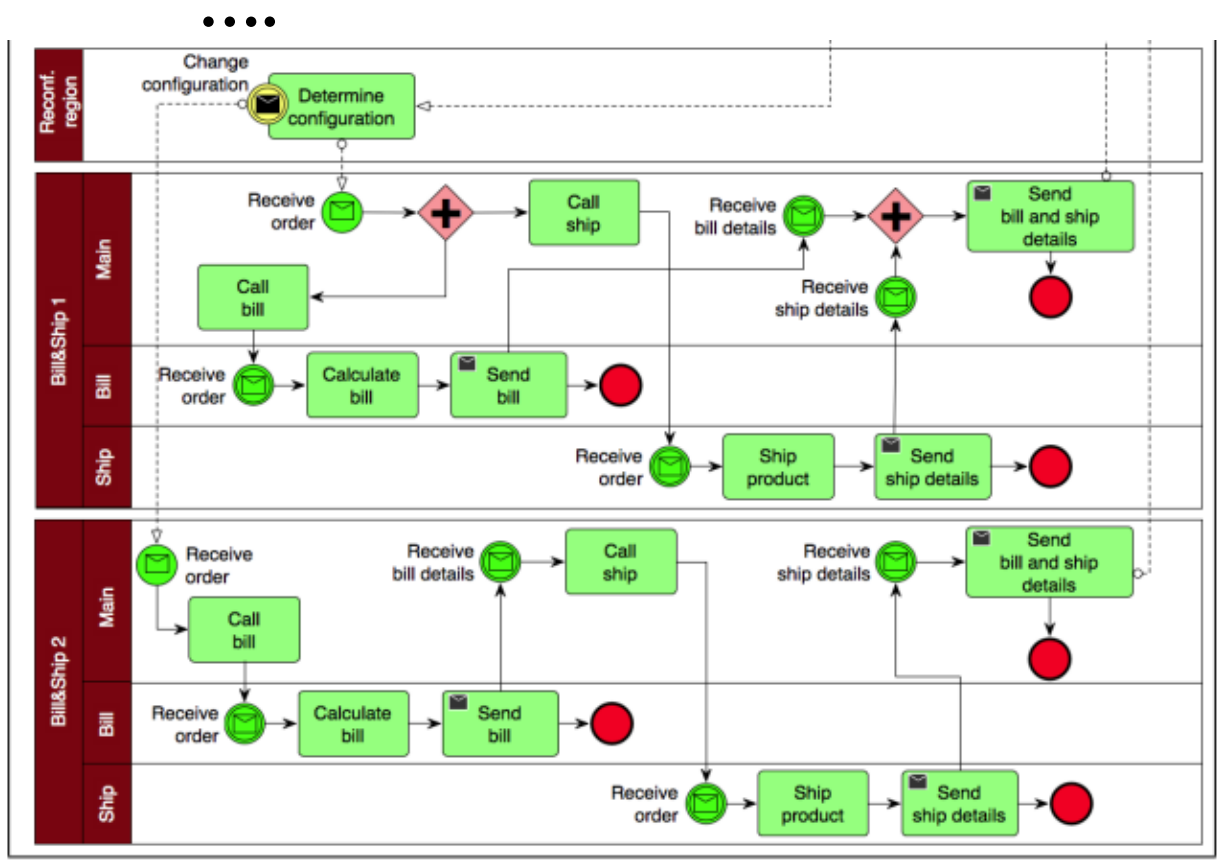

Fig. 4 The modified part of new configuration BPMN diagram

## 3.2 TM Static Model

Fig. 5 shows the corresponding TM static model. The figure can be described as follows. First, a customer request arrives at the office (circle 1). It is sent to the order generator, (2) where an order is created (3) and sent to the office, where it is processed (4). If it is OK, a credit check is created (5) that flows to the credit check unit, (6) where the result of checking is created (7) and flows to the office (8). If the result is not OK (9), a rejection is sent to the customer (10 and 11). If the result is OK, (12) an inventory check is created and sent to inventory (13). The result of the inventory check is sent to the office (14 and 15), where it is processed. If the result is negative, a rejection is sent to the customer (16 and 17). If the inventory result is positive (18), a Bill&Ship request is sent to Bill&Ship main (19). There, requests for a call bill (20) and call ship (21) are sent to the billing and shipping units. When the billing details (22) and shipping details are received (22 and 23), an invoice with billing and shipping details is created (24) and flows to the archive (25). When an acknowledgement is received from the archive (26), a notification of success is sent to the customer (27 and 28).

## 3.3 TM Events Model

A TM event is defined as a subdiagram of the static diagram (called a region of the event) plus time. Fig. 6 shows a sample of two events: *calculating the billing details* and *calculating the shipping details*. Accordingly, we can specify events on the static TM diagram assuming that regions represent events. Each generic event can be converted to a generic event; however, models usually specify larger events.

The set of events defined over the static description can be listed as follows.
Event 1 ($E_1$): A customer request reaches the office.
Event 2 ($E_2$): The request is sent to the order generator, where it is processed.
Event 3 ($E_3$): The order generator creates an order.
Event 4 ($E_4$): The order flows to the office workflow, where the order is examined.
Event 5 ($E_5$): If the order is OK, a request for a credit check is created.
Event 6 ($E_6$): The request for a credit check goes to the credit check department.
Event 7 ($E_7$): The credit check department processes the request and creates OK if it is valid.

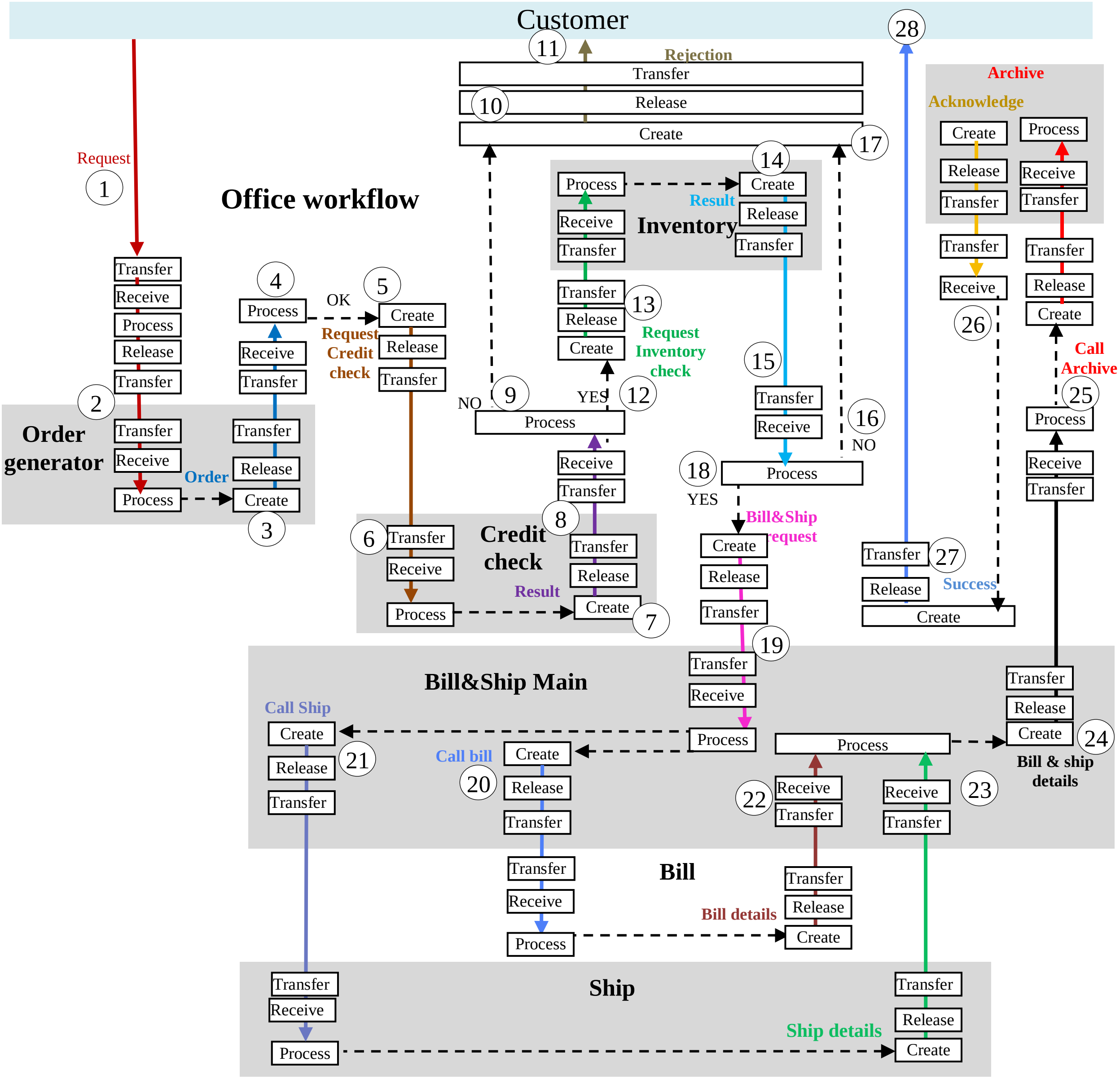

Fig. 5 The TM static model of the case study

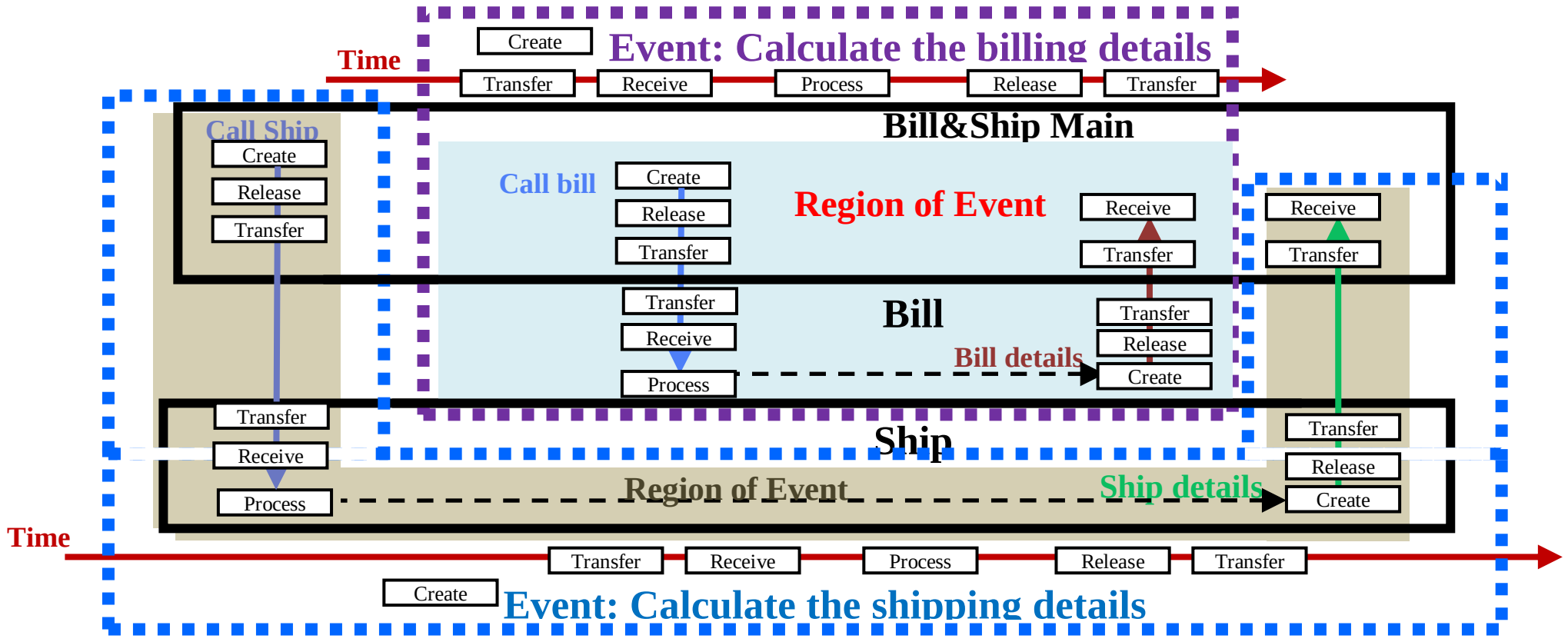

Fig. 6 Two events of the case study

To save space, we will not list all nineteen events that Fig. 7 shows.

From the chronology of these events, we can construct the behavioral model of the system of handling product orders from customers, as Fig. 8 shows.

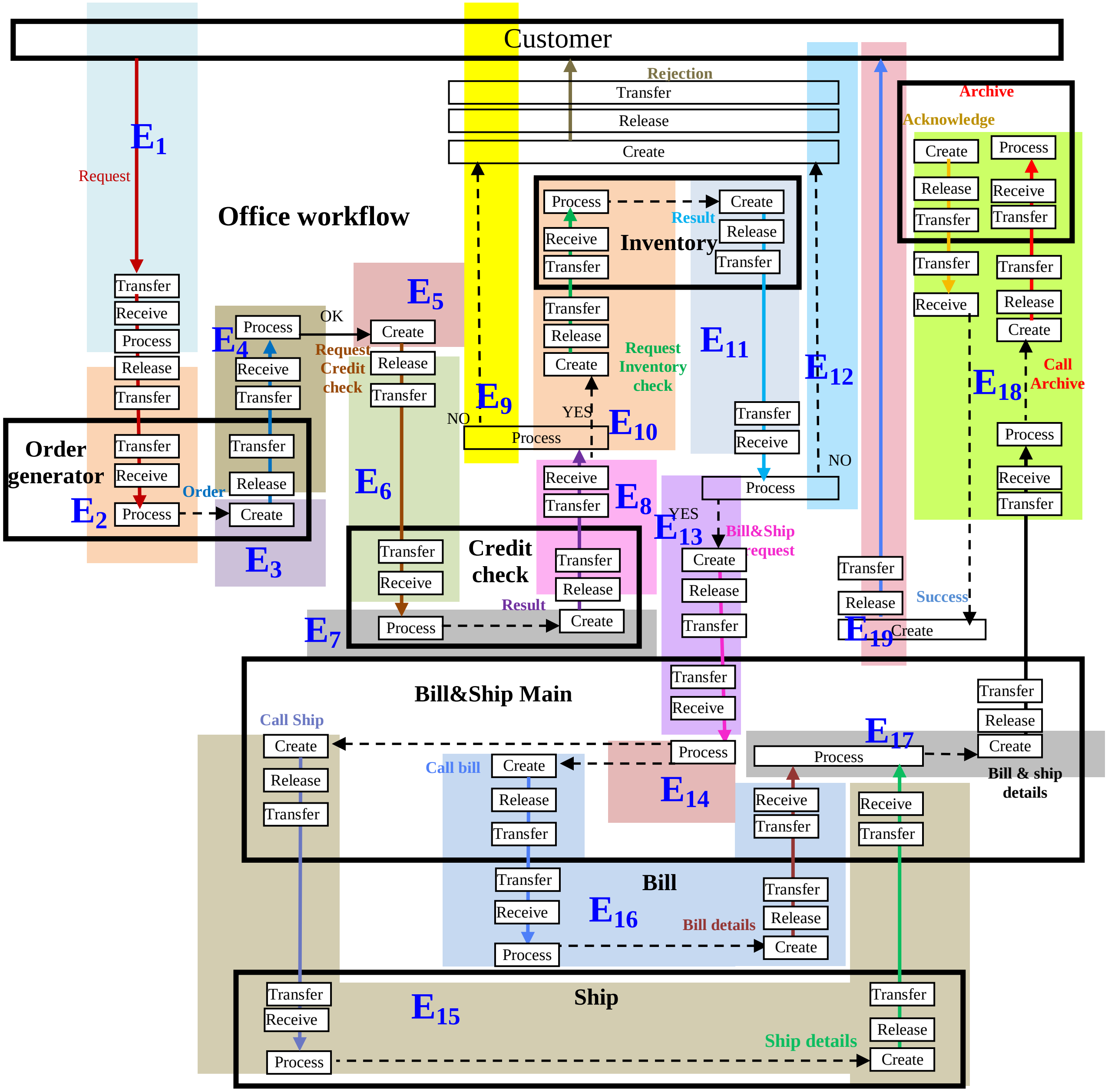

Fig. 7 TM events model of the case study

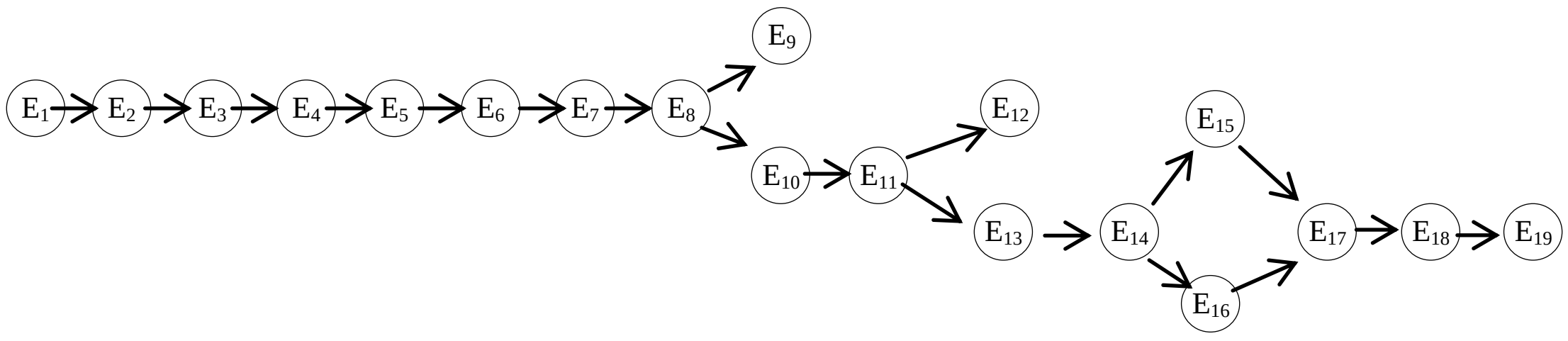

Fig. 8 TM behavioral model of the case study

### 3.4 Reconfiguration

Zhou's [7] reconfiguration of the order of billing and shipping activities involves moving the billing activity to occur before the shipping activity and keeping the old and new configuration processes simultaneously available. From the TM modeling prospect, it is clear that the required reconfiguration aims at change in the behavioral model. This change can be accomplished without changing the original static TM description. In contrast, Zhou's BPMN model involves a single static-level description; therefore, it is necessary to change the original BPMN diagram. In the TM, we can introduce two additional events as follows:

$E_{20}$: Configuration of behavior 1, where the billing activity and the shipping activity are simultaneously activated.
$E_{21}$: Configuration of behavior 1, where the billing activity occurs before the shipping activity and bookkeeping.

Accordingly, the behavioral model is structured as Fig. 9 shows. The change involves the control system, which activates various modules of events. Imagine Zhou's configuration problem involves n>2 series of alternatives. This situation requires an extensive additional modification to the original schema. A TM events-level solution is a more effective solution.

## 4. A Glimpse of the Change Notion

The remaining part of this paper will focus on exploring the notion of change. Because the philosophical field of the topic is very broad, we will concentrate on a specific track of study, namely Zeno paradoxes.

Generally, "change" is used to refer to an object changing its ordinary properties over time. An object undergoes change whenever it gains or loses some of its properties. According to Mortensen [8], "Change is so pervasive in our lives that it almost defeats description and analysis." Change is problematic because it requires something becoming something out of something that it is not. Also, as change is a fluid process, how can we accurately determine when it has happened or to what extent? [4].

In ancient Greece, Heraclitus was famous for his insistence that "the only thing that is constant is change." Heraclitus claimed that everything changes all the time. Aristotle claimed that "time is the measure of change" and that "there is no time apart from change…." (*Physics*) of things. According to Aristotle, "time is not change [itself]" because a change "may be faster or slower, but not time…." (*Physics*). In a modern context, Einstein's theory of relativity implies a moving clock can tick more quickly or slowly than another clock, but time itself isn't faster or slower [9].

Zeno argued that change cannot exist; it is all an illusion. The argument in this context is that things cannot exist and not exist simultaneously. Reality is an unchanging unity. Zeno developed a series of paradoxes to demonstrate logically that change is an illusion. Motion, for example, is an illusion. To reach a destination, one must first reach a halfway point. When one reaches that halfway point, they have yet another one to reach, and this process is logically infinite. Movement does exist, but how can we possibly determine exactly when and how it happens without involving Zeno-type paradoxes [4]?

In the next section, we apply one of the Zeno-type paradoxes in the context of TM where space is conceptualized as thimacs. Accordingly, we will visualize "space units" as thimacs with their five action potentialities (no speculation about the nature of these thimacs).

A thimac can be viewed as having interior (create, process, and receive (accept)) and boundary (release, transfer, and receive (arrive)) posts. A moving thing may reach the boundary of this space thimac (arrive) but not necessarily "enter" it (accepted). Some factor (e.g., movement) leads the arriving thing to be ejected to the outside (transfer: output) of the space thimac. Fig. 10 illustrates this scenario. Therefore, we have two disjoint cases: inside the space thimac, where settlement and continuity occurs, and the thimac boundary posts, where moving things collide with the space thimac but never succeed in penetrating to the inside. Accordingly, a motion is possible among these space thimacs similar to a child walking inside a container filled with plastic balls.

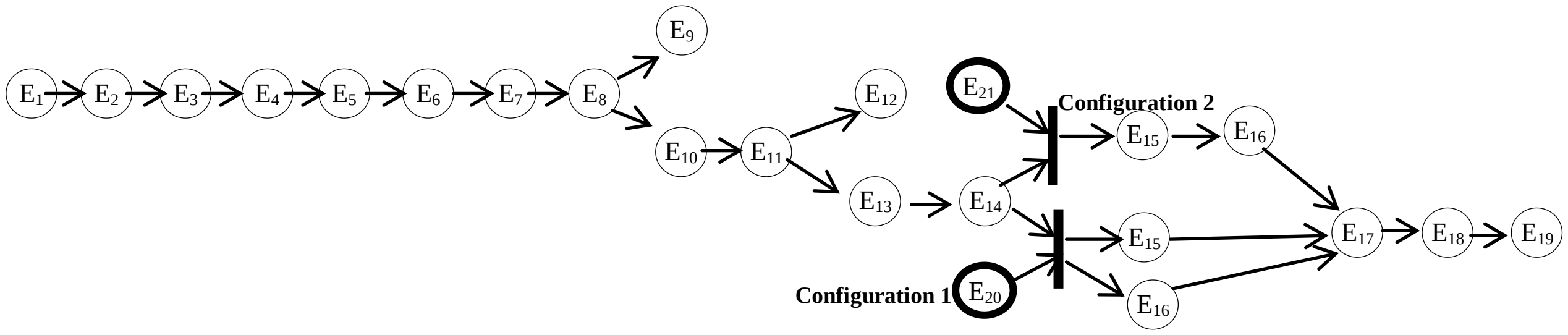

Fig. 9 TM new behavioral model of the case study

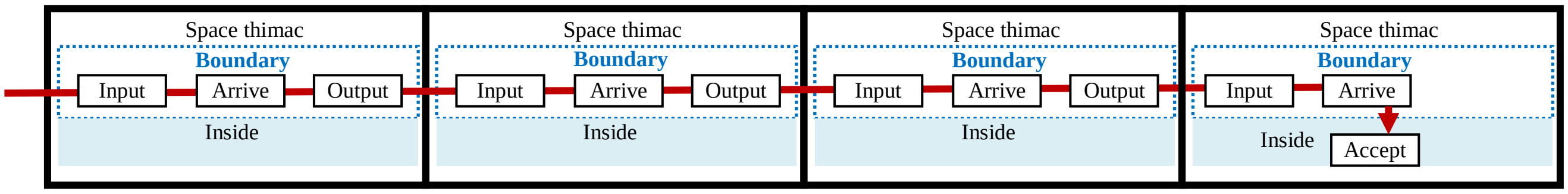

Fig. 10 A thing flows across space thimacs where each thimac does not accept it until it loses its movement energy and settles in the last thimac

However, we assume that such bouncing of things among space thimacs is applied to things in motion. Stable things can settle inside these space thimacs. The following section elaborates on such speculative analysis and applies it to Zeno's puzzle of an arrow's movement. We explore the notion of change by analyzing Zeno's paradox of motion using the TM model. A paradox is a proclamation that holds conflicting concepts. The discussion is pure speculation and may contribute to logical analysis of the concept of *space*.

## 5. Change and Zeno's Paradox

One philosophical approach claims that space (and time) exist always and everywhere regardless of what else exists and that space (and time) provide a container within which matter exists and moves independently of the container [9]. Newton argued that space is an absolute entity and that everything moves in relation to it. This concept led to the distinction between "relative" and "absolute" motion. Leibniz maintained that space is the spatial relations between things. Space cannot exist independently of the things it connects. If nothing existed, then space would not exist. This statement implies that there is no container; space is only a set of relationships among existing physical material, and time is a set of relationships among the events of that material [9].

Current thinking is that space is quantized; therefore, when we move across space, we are actually jumping from small locations to other small locations [10]. As Cham and Whiteson [10] stated, in this view, space is a network of connected nodes, like the stations in a subway system. Each node represents a location, and the connections between nodes represent the relationships between these locations (i.e., which one is next to which other one).

These nodes of space can be empty and still exist. A field just means there is a number, or a value, associated with every point in that space. In this view, particles are just excited states of these fields [10].

### 5.2 Space and Motion

Zeno's paradox under consideration deals with the problem of how to account for change and continuity together.

We cannot assert that an arrow is actually moving after it has been shot from a bow because the arrow needs to be at a certain place at each point of time, which by definition cannot contain any duration at all [11]. The arrow is not moving because all of its trajectories consist of a series of these moments, and at each moment, it is not moving. So if it is not moving at one moment, then it is not moving at all [11]. According to Hongladarom [11],

> One might, for example, argue against Zeno that points of time containing no duration at all do not actually exist and what do exist are only chunks of time which contain a length of time however small. Hence there is not such a thing as a point in the line of time, and what does exist in the line are smaller sections of the line which can be divided and further divided, but no absolute point can be reached. The arrow, then, moves in these smaller chunks of time, and since these chunks are not points the arrow can move within those chunks. This argument does not seem to work, however, because one would then need to find an account of how the arrow moves from the beginning of the chunk to the end.

The basic problem, according to Hongladarom [11] is the simultaneous existence of continuity and change in the case of the moving arrow. Mortensen [8] described the arrow puzzle as follows:

> An arrow in flight could not really be moving because at any given instant it would be at a place identical with itself (and not another place); something at just one (self-identical) place could not be described as moving, and an arrow which is motionless at every instant in a temporal interval must be motionless throughout the interval.

Following the view that *space is a network of connected nodes*, we propose that these nodes are thimacs. Therefore, we have a net of thimacs, as Fig. 11 shows, illustrated as connected machines in two-dimensional background. Two connected machines denote connectedness that permits flow of things between the thimacs.

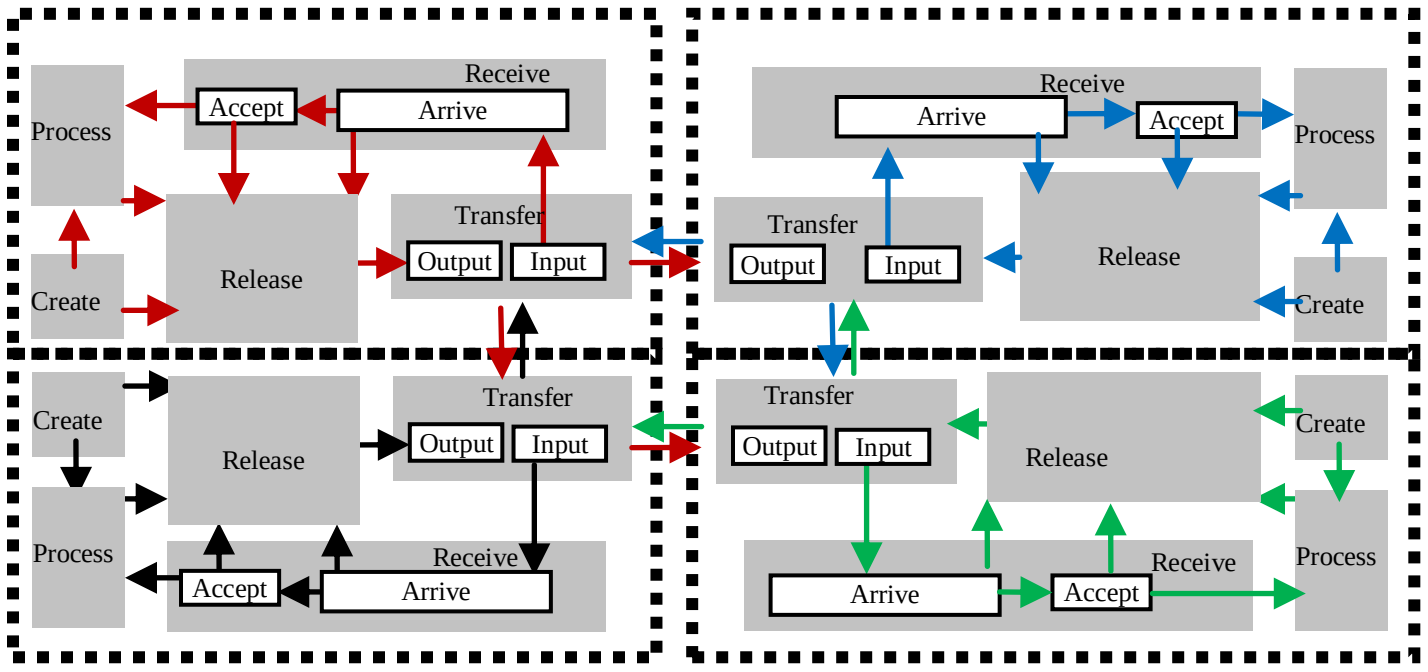

Fig. 11 Thinging machines adjacent to each other

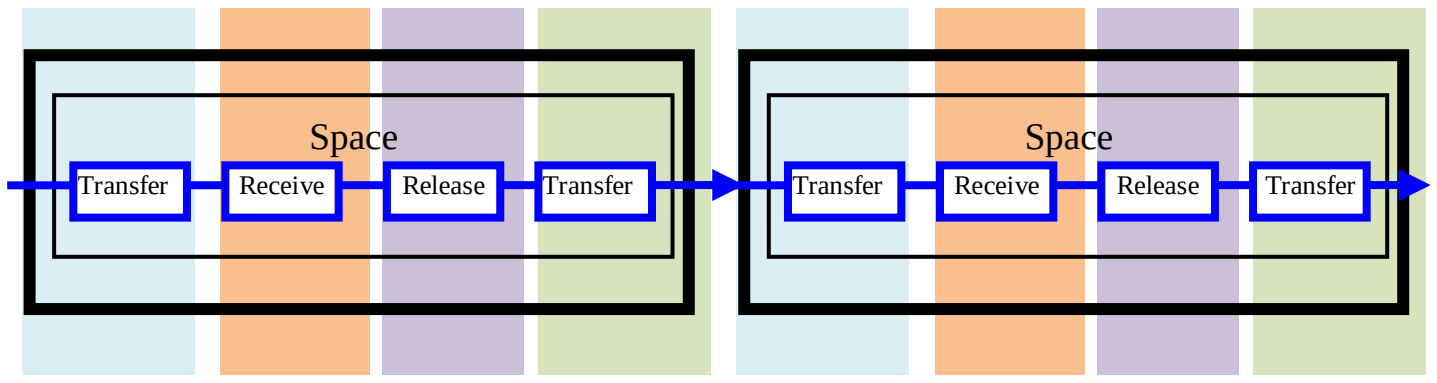

Fig. 12 Propagation of change between machines in terms of sequence of events, assuming that each machine represents a unit of space

Events (changes) propagate between the two connected thimacs through the transfer stages that form their "borders." For example, Fig. 12 shows the propagation of change (events) between two machines in terms of sequence of events, assuming that each machine represents a unit of space. Next, we distinguish between two types of phenomena in a TM: states and progressions. The idea is that progression is unstable change that does not need to be the thing to be in a fixed place. For example, suppose that a ball hit a player and bounced away. In this case, we can say, using TM terminology, that the ball has been transferred, arrived to the player, then transferred away without being received by the player. Similarly, an arrow can "touch" the space thimac without being received into the corresponding space.

A TM stage can be viewed as either a state or progression (typically called PROCESS). When understood as a state, a stage is a complete potentiality ready to be actualized. Understood as a progression, it is an accumulated PROCESS. A transfer event between two thimacs is a pure progression stage. For example, when an arrow in the Zeno puzzle is in the transfer event between two space thimacs, this does not mean that it is located in the transfer stage. *Transfer itself is a pure change*; therefore, change cannot be captured as a stable phenomenon. In contrast, a process event is a change and a state. For example, changing the color of a car from blue to white involves the process of "whiting" the car and the result: the car is now a white car. Transferring an arrow between two adjacent spaces (e.g., through a door between them) involves movement but no *location* where the arrow can be in the transfer state. Creation also involves a location, e.g., the arrow is created and ready to be released or processed.

In TM, a change happens *in* the model when a generic event occurs, e.g., creation, processing, releasing, transferring, and receiving. A generic event can be understood in term of states and/or a progression

- When understood as a *state* of a thing (such as being created), as in physics, a state of matter is one of the distinct forms in which matter can exist, i.e., solid, liquid, gas, or plasma. Accordingly, states of things in a TM machine are created, processed, released, transferred, and received. For example, the state *created* indicates existence, which may be declared initially as in "there is" or appears as a result or effect of processing, e.g., salt (NaCl) appears from processing its chemical components, sodium (Na) and chlorine (Cl).

- Understood as a *progression*, the event is the rising stretch of flux. A thing is subjected to alteration to reach the state of being processed, e.g., a car is exposed to damaging before being labeled as a damaged car; a patient needs to be anesthetized to be in the state of unconsciousness.

Accordingly, we can redefine a thinging machine to emphasize the internal states and progression actions, as Fig. 13 shows. The machine is described as follows:

- Create: a thing is created initially (given) by a previous process in the machine (circle 1), e.g., processing Na and Cl creates NaCl. As a result of the creation event, a thing is in the created state. A state will be denoted by a thick horizontal bar, as the figure shows. There may be a queue of created things in this state. The thing in the creation state may flow to either the process stage or the releasing stage.

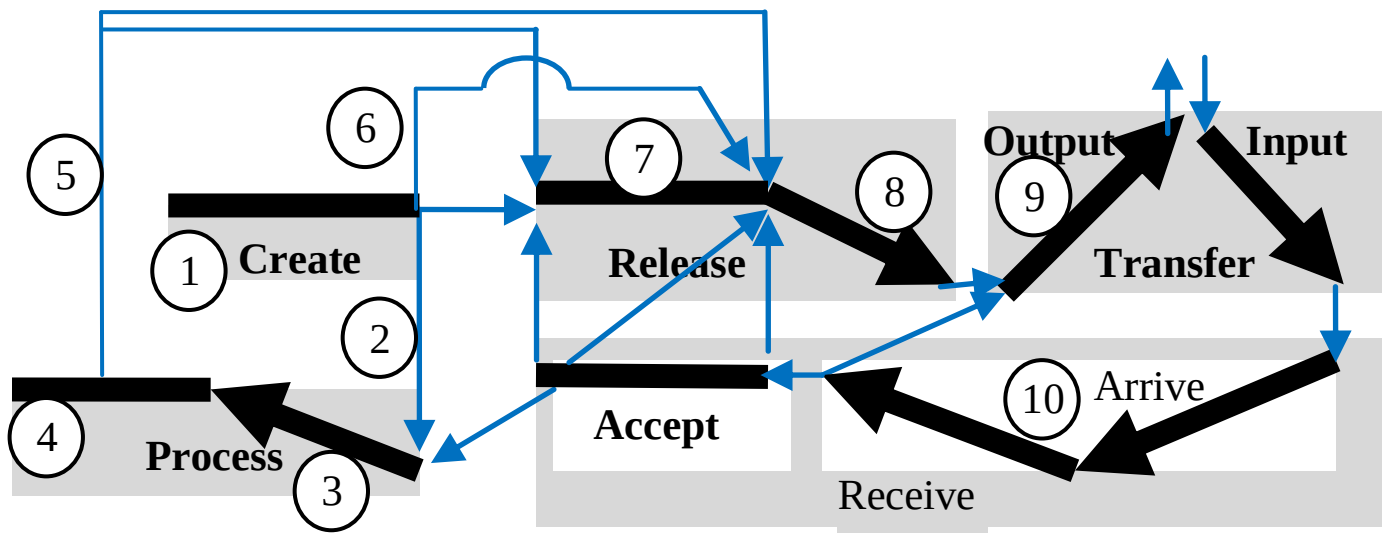

Fig. 13 States and progressions in a thinging machine

- If a created thing flows to the process stage (2), then the first phase of the processing includes a progression to change the thing (3). This progressing is denoted by an upward thick arrow, as the figure shows. Thus, the thing first starts changing, half changing, etc. until it reaches the changed (processed) state, which is indicated by the horizontal thick bar (4). It may stay in this state for a while, or it may flow to either the released state or progression (released immediately) (4).

- If a created thing flows to the release stage (6), it may stay in the released state (7), waiting for a while to be actually released, or it may be subjected to releasing (progression) immediately (8). This progression event is denoted by a downward thick arrow. It involves continuous motion of a thing to the thimac's boundary, or "periphery."

- A good example of the release state is an output buffer filled with data waiting to be released to the outside. The progression starts with bits leaving the buffer and racing to reach the output port. This "race" is a progression.

- Transfer – Output (9): This is a progression represented by "jumping" to the outside. The jump itself is not a state and involves time. No stable state exists between racing to the edge (release progression) and jumping to the outside (output progression). It is a continuous change.

- Arrival (10): This is a progression of a thing reaching a machine's boundary from the outside. At this point, the thing is not in the machine but only arriving and interacting with this boundary to be accepted or castoff to the outside. It is a progression analogous to an ocean wave reaching upward by a push of air to reach a high point and then collapse downward. The wave is never in a recognizable state. Similarly, at the thing's arrival, it evolves and collapses in being judged to be accepted or pushed back to the outside (we denote this progression with two thick arrows in a reverse form of a wave).

A thing being subjected to progression means that it is never "motionless" in the involved stage. Fig. 14 shows an analogy for this scenario, in which a bed frame is moved through various floors without actually being on any one floor before reaching its destination. Another analogy is a traveler who arrives at several transit countries before reaching their destination.

Applying this concept to Zeno's arrow, the moving arrow is never in two successive place units at one time, and no diverse states of the arrow exist within one time. Fig. 15 illustrates this situation in which the arrow flows to input, arrival, and output. See also Fig. 16. The arrow "squeezes" through the space units, bouncing back at the entrances, as Fig. 17 shows. As soon as it loses its motion, its momentum is "absorbed" inside a space unit.

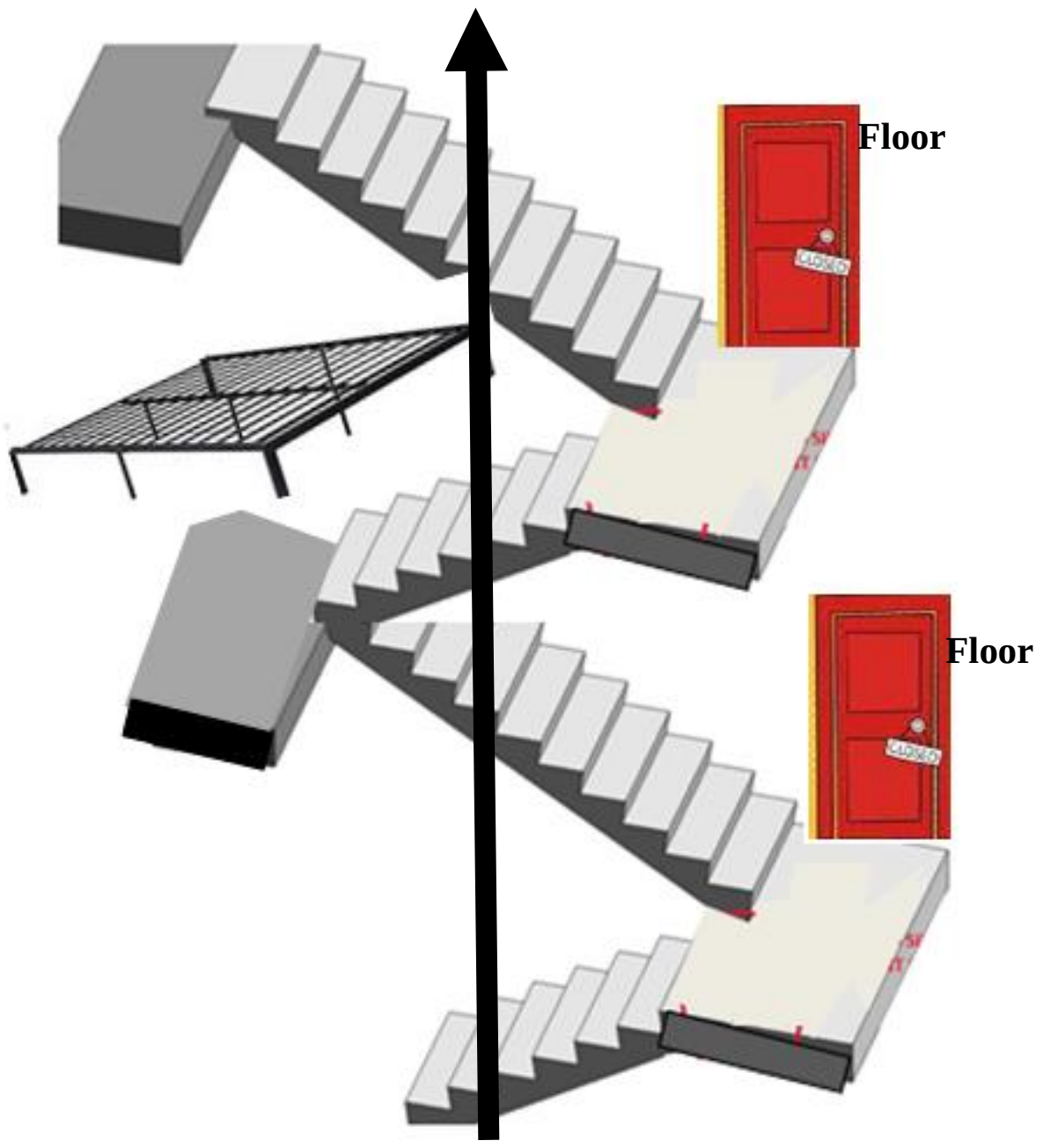

Fig. 14 A bed frame moves between floors; however, at no particular point in time it is at a place that is part of any floor.

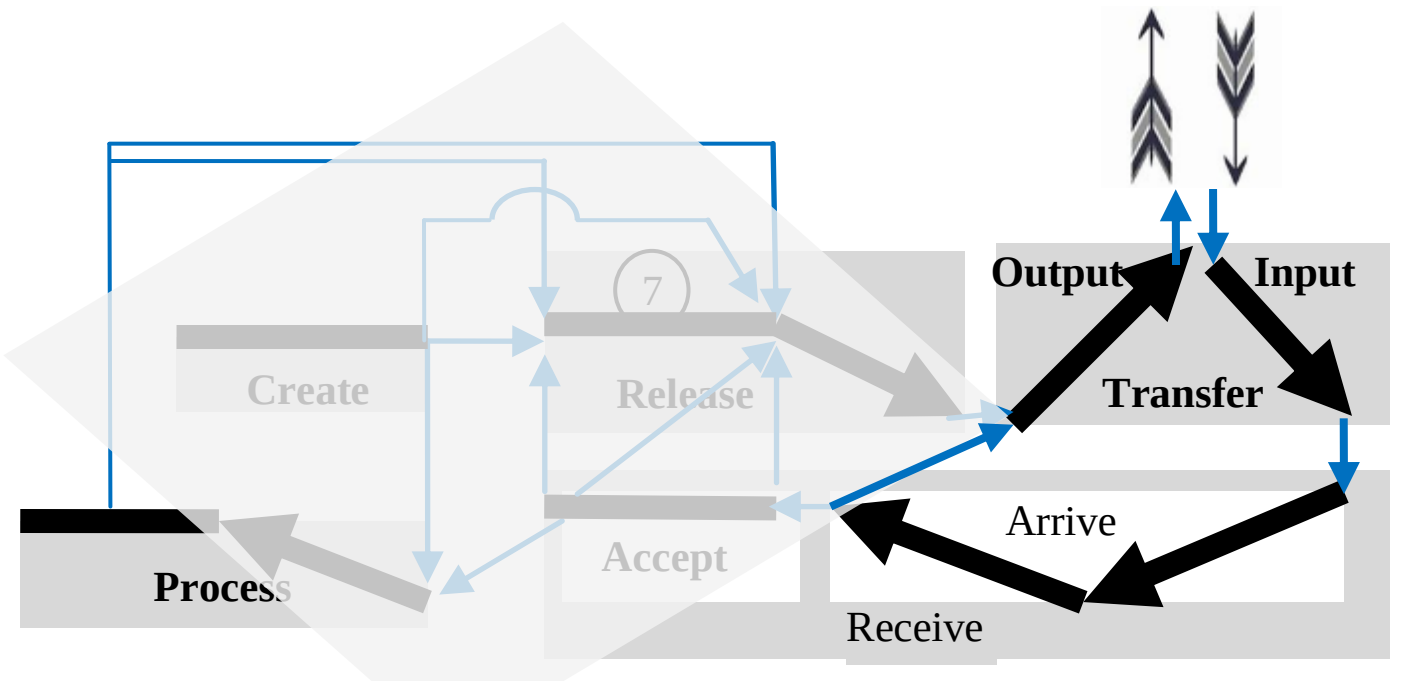

Fig. 15 The projection of the arrow in a thinging machine

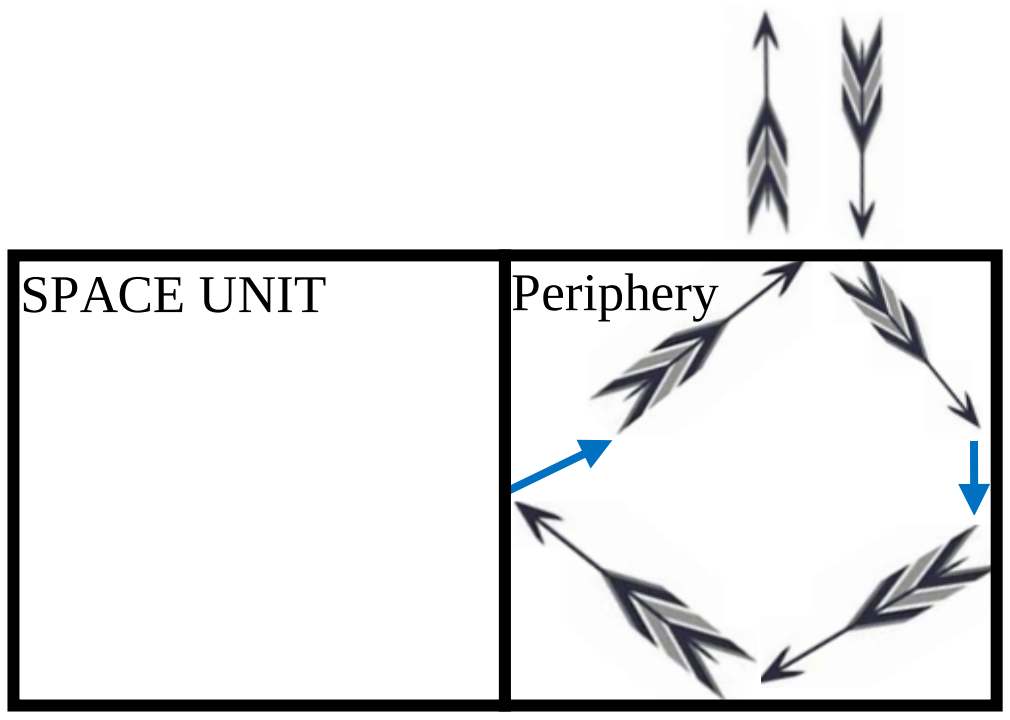

Fig. 16 The arrow does not settle in a thinging machine

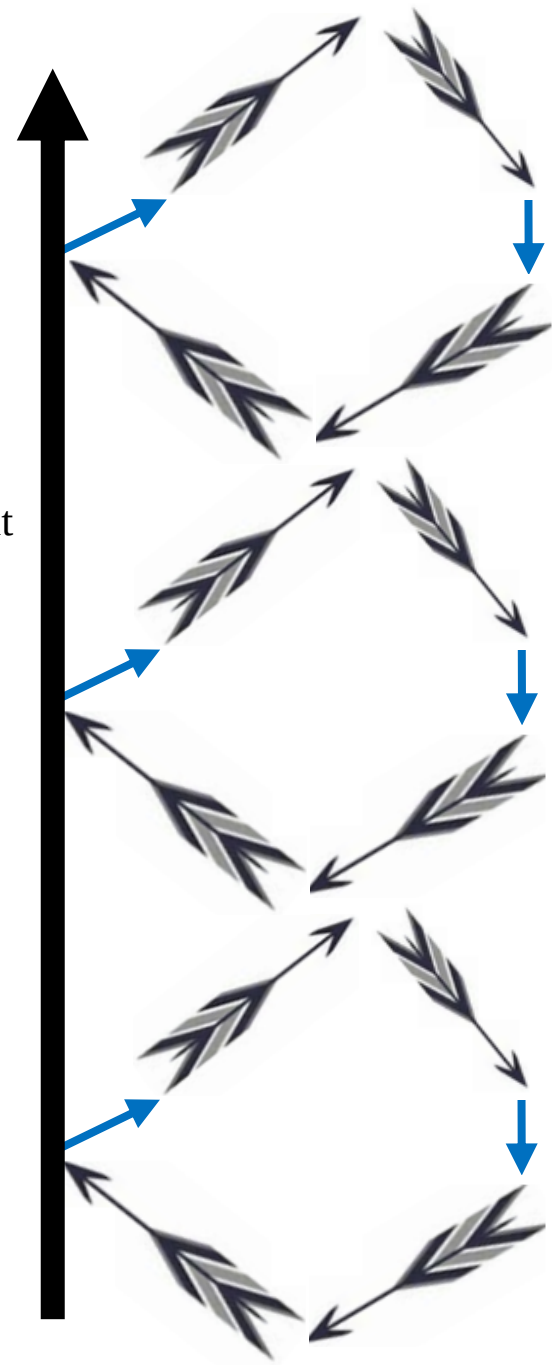

Fig. 17 The arrow "squeezes" through the space units, bouncing back at the entrances

## 6. Conclusion

In this paper and several previous papers, we have explored the TM model and its application in several notions, such as change, event, and systems behavior. This type of study enhances the software and systems engineering community's philosophical foundations. In this article, we aimed to reflect on philosophical concepts, specifically the concept of change, and attempted to introduce conceptual modeling in the philosophical context. We started by giving a sample TM application in software engineering in the form of a conceptual model of a business orders system. The example involves modification of an ordering conceptual schema as an instance of requirements change in the original description of the system. This case study of change in a system led to application of the same modeling tool, the TM model, of change in the philosophical sense, as in the problem of an arrow's movement in Zeno's paradox. Although the general idea of dividing space into connected nodes is not a new concept, applying thimacs as space units seems to introduce a logical explanation for the movement discussed in Zeno's puzzles. This direction of study needs further scrutiny and examination, but it seems to be a promising exercise that ties together problems in several fields of study.

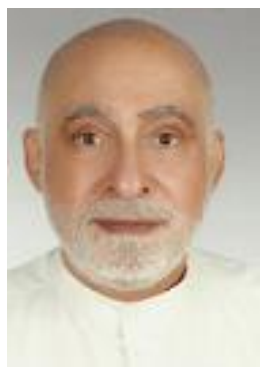

**Sabah S. Al-Fedaghi** is an associate professor in the Department of Computer Engineering at Kuwait University. He holds an MS and a PhD from the Department of Electrical Engineering and Computer Science, Northwestern University, Evanston, Illinois, and a BS from Arizona State University. He has published many journal articles and papers in conferences on software engineering, database systems, information ethics, privacy, and security. He headed the Electrical and Computer Engineering Department (1991–1994) and the Computer Engineering Department (2000–2007). He previously worked as a programmer at the Kuwait Oil Company. Dr. Al-Fedaghi has retired from the services of Kuwait University on June 2021. He is currently (Fall 2021/2022) seconded to teach in the department of computer engineering, Kuwait University.